\documentclass[11pt,a4paper]{article}
\usepackage{amssymb,amsmath,float,tabularx,amsfonts,booktabs}
\usepackage{graphicx}
\usepackage{color}
\usepackage{setspace}
\usepackage[normalem]{ulem}
\usepackage[toc,page]{appendix}
\usepackage{authblk}
\usepackage[backend=biber,sorting=none]{biblatex}
\usepackage{algorithm}
\usepackage{algpseudocode}
\usepackage{titlesec}
\usepackage{fdsymbol}
\usepackage[a4paper,margin=3cm]{geometry}
\usepackage[table]{xcolor}
\usepackage{multirow} 
\usepackage{hyperref}
\hypersetup{colorlinks,allcolors=blue}
\makeatletter
\usepackage{nicematrix}
\newcommand{\multiline}[1]{%
    \begin{tabularx}{\dimexpr\linewidth-\ALG@thistlm}[t]{@{}X@{}}
        #1
    \end{tabularx}
}

\begin{document}
\title{Spatio-temporal agent-based modelling of malaria}
\author[1$,2\clubsuit$]{Camelia R. Walker }
\author[1$\clubsuit$]{Md Nurul Anwar}
\author[3,4,5]{Leandra Bräuninger}
\author[6,7]{Jack Richards}
\author[8]{Ricardo Ataide}
\author[9]{Ngo Duc Thang}
\author[9]{Nguyen Xuan Thang}
\author[6]{Sara Canavati}
\author[1]{Jennifer A. Flegg}

\affil[1]{School of Mathematics and Statistics, The University of Melbourne, Parkville, Australia}

\affil[2]{Melbourne School of Population and Global Health, The University of Melbourne, Parkville, Australia}
\affil[3]{Nuffield Department of Clinical Medicine, University of Oxford,  United Kingdom}
\affil[4]{Department of Statistical Science University College London, United Kingdom}
\affil[5]{The Alan Turing Institute, London, United Kingdom }
\affil[6]{Burnet Institute, Australia}
\affil[7]{ZiP Diagnostics, Australia}
\affil[8]{Infection and Global Health Division, The Walter and Eliza Hall Institute, Parkville, Australia}
\affil[9]{National Institute of Malariology, Parasitology and Entomology, Vietnam}

\date{}                    
\setcounter{Maxaffil}{0}
\renewcommand\Affilfont{\itshape\small}
\maketitle
$\clubsuit$\ \small{These authors contributed equally to this work}

\maketitle
\begin{abstract}
    \textit{Plasmodium falciparum} is responsible for the majority of malaria morbidity and mortality each year. Malaria transmission rates vary by location and time of year due to climate and environmental conditions. We show the impact of these factors by developing a stochastic spatiotemporal agent-based malaria model that captures the impact of spatially distributed interventions on malaria transmission. Our model uses spatiotemporal estimates of mosquito climatic suitability and household location data to model the interaction between human and mosquito agents. We apply our model to investigate how strategies for distributing interventions to households in Vietnam impact the disease burden. Our study shows that providing some level of protection to a wide range of households reduces malaria prevalence more compared to providing a strong level of protection to a limited number of households. 
\end{abstract}

\section{Introduction}
Malaria is a mosquito-borne disease that causes significant morbidity and mortality around the world, especially in the tropics. Although malaria is treatable and curable, it is still one of the leading causes of death in large parts of Africa and Asia. According to the WHO Malaria Report, in 2023, an estimated 263 million cases of malaria occurred globally, with the Western Pacific Region accounting for an estimated 1.74 million malaria cases \cite{world2024world}. An estimated 448 malaria cases occurred in Vietnam in 2023, where the central region of Vietnam continues to be a malaria hotspot, accounting for the majority of cases reported in the country \cite{world2024world,khanh2025unprecedented}. However, the overall trend in malaria cases in Vietnam is decreasing and is nearing elimination. \\

 Many countries affected by malaria have limited resources available for the treatment and interventions needed to eliminate the disease. This presents an optimal decision problem: to determine the intervention strategy that minimises malaria prevalence while constrained by a fixed budget. An intervention strategy can be considered in terms of the type of intervention that is delivered and how that intervention is implemented.  To determine the relative benefit of an intervention strategy, their varying costs and effects on malaria transmission dynamics should be considered; for example, some primarily function by killing mosquitoes, while others act by preventing mosquito bites. When considering how an intervention is targeted, it is also important to consider whether environmental conditions favour malaria transmission and the impact of household size on transmission.\\
 
 In Vietnam, there are a variety of types of interventions currently in use, including long-lasting insecticide nets (LLINs) and indoor residual spraying (IRS) \cite{van2023anopheles}. LLINs are bed nets that contain a WHO-recommended insecticide and form a physical and chemical barrier against mosquitoes when used and protect people at risk. They are one of the most effective vector intervention measures, having reduced the incidence of malaria by 50\% in sub-Saharan Africa, in a region that accounts for more than 90\% of global cases \cite{unicef2022fighting}. These chemically sprayed bed nets can last even after 20 washes. However, the improper use and disposal of LLINs have become an issue of concern with regard to the environmental and ecological impact \cite{santos2021unsustainability}. IRS targets mosquitoes that rest within the home by applying effective residual insecticides to the walls of the home where mosquitoes are known to rest \cite{pluess2010indoor}. A study conducted in Zambia showed that IRS contributed to 25\% of the decline in parasite prevalence during the first 6 months of the rainy season \cite{zhou2022effectiveness}. Although both LLINs and IRS have shown substantial efficacy, the growing prevalence of insecticide resistance in many endemic regions poses a significant challenge to their sustained effectiveness \cite{riveron2018insecticide,world2024world}.\\

Various environmental factors can affect the density and movements of \textit{Anopheles} mosquitoes, the genus capable of transmitting malaria to humans, at a given location. Temperature, humidity, rainfall, and the presence of bodies of water all impact local mosquito density \cite{Endo:2018, Huang:2011}. In addition, these environmental factors exhibit seasonal variation. The Malaria Atlas Project \cite{Hay:2006} has produced global maps of suitability for \textit{P. falciparum} malaria based on temperature, and Brown \textit{et al.} \cite{brown2024global} have extended this work by incorporating additional seasonal factors, namely humidity and rainfall.
The transmission of malaria within a geographical location is driven by both environmental conditions and the structure of the human population. Individuals living in areas where malaria is highly prevalent or those who engage in an activity that brings them into frequent contact with infected mosquitoes (e.g., forest work) are more likely to be exposed to the parasite than the average person \cite{anwar2024mathematical}. Malaria infections are spatially heterogeneous in remote forested regions in Vietnam, and risk can vary substantially within and between villages \cite{bannister2018micro}. As mosquitoes can only travel a limited distance and some bite multiple times in a day, household clustering could be an important factor in the dynamics of malaria. \\

The complexity of malaria transmission dynamics necessitates the use of advanced modelling techniques to capture the interactions between environmental factors, human behaviour, and vector populations.  Mathematical modelling has proven instrumental in providing crucial insights into infectious disease dynamics \cite{keeling2011modeling}. In particular, it plays a key role in informing public health policy by assessing the impacts of various intervention strategies \cite{huppert2013mathematical}. 
Unlike classical compartmental models, which typically aggregate individuals into homogeneous groups, agent-based modelling (ABM) offers significant advantages for understanding malaria transmission due to its ability to simulate individual-level interactions and incorporate spatial heterogeneity, such as varied exposure to mosquito bites and uneven implementation of interventions \cite{Smith:2018}.\\

In this paper, we present a spatiotemporal agent-based model that accounts for spatial heterogeneity in bite exposure based on malaria suitability estimates \cite{brown2024global} and the impact of vector interventions on a household scale in Vietnam. Our objective in this work is to investigate the impact of the spatial distribution of additional interventions on \textit{P. falciparum} malaria burden. The article is structured as follows. In Section 2, we introduce the input data for humans and mosquitoes in the study region. We describe the model development and additional intervention strategies in Section 3. We provide results from the ABM in Section 4 before making our concluding remarks in Section 5.

\section{Data}

\subsection{Malaria intervention and infection data from Vietnam}
To investigate how spatial differences in interventions affect \textit{P. falciparum} malaria burden, we model spatiotemporal malaria dynamics, considering the example of \textit{P. falciparum} in Vietnam. We utilise human data from a household survey from 2017 in five communes in Dak Nong (Dac O, Bu Gia Map, and Dac Nhau) and Binh Phuoc (Quang Truc and Dac Ngo) provinces in Vietnam. The survey contains two data sets: (1) a GPS data set consisting of households with location, household ID, household size, individual ID, commune ID, and malaria interventions in use; and (2) a case dataset consisting of individuals who tested positive for malaria with household location, household ID, and individual ID. Among household variables, infection status and interventions vary over time and are dynamically modelled through the ABM (see Section \ref{ABM}).\\

In the case dataset, we found that of the 91 malaria cases (42 were infected with \textit{P. falciparum}), 33 have household IDs that do not match any ID in the GPS data set. As the case data set contains no variable for the household size, we cannot determine the household size of those individuals. Instead, we assign those individuals a household size by sampling from the observed distribution from the GPS dataset. After combining the GPS data with the case data set and filtering out duplicate households, there remain a total of 10,380 unique households made up of a total of 45,466 individuals. There are also 103 households (1\%) for which no information on intervention measures is available, and we assume that there are no interventions in these households. \\

\begin{figure}[!ht]
\centering
\includegraphics[width=\textwidth]{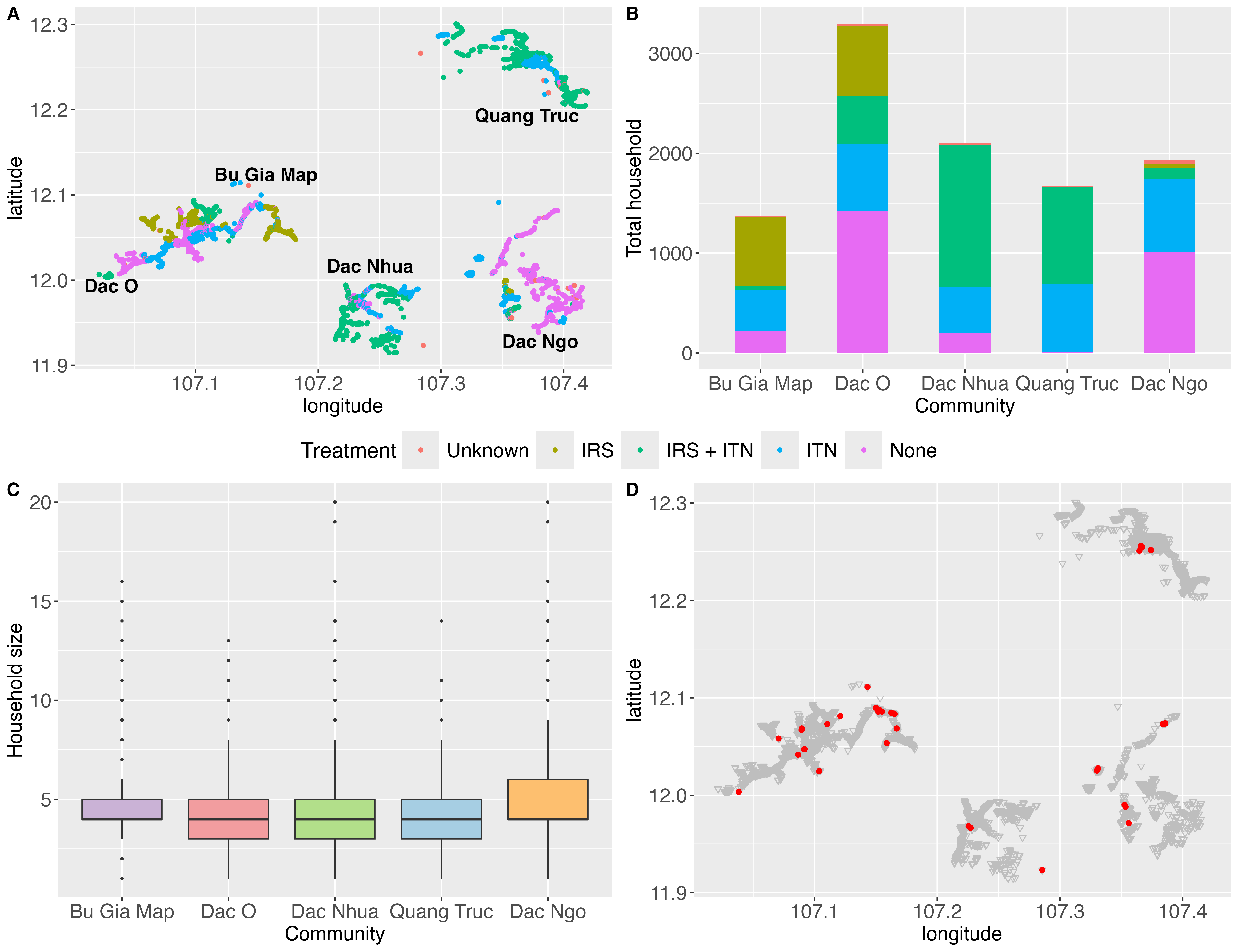}
\caption{Household data and malaria interventions. \textbf{A:} Households employing the different intervention strategies. The households are colour-coded by their current intervention strategy \textbf{B:} The counts of households employing the different intervention strategies per community. \textbf{C:} Distribution of household size within each community and   \textbf{D:} The location of infected individuals.}
\label{fig:all_data_hum}
\end{figure}

Figure \ref{fig:all_data_hum}A shows the location of the households and the interventions they used. Figure \ref{fig:all_data_hum}B illustrates the breakdown of community interventions. The Dac O commune has the largest population of the five communes, comprising 31.8\% of the households. Dac O also contains the highest proportion of households without intervention measures (27.5\%). Of 10,380 total households, 3019 households (29.1\%) use IRS and LLINs, 2953 households use LLINs only (28.4\%), and 1447 households use IRS only (13.9\%). Figure \ref{fig:all_data_hum}C shows the distribution of household size, while Figure \ref{fig:all_data_hum}D shows the location of the individuals who tested positive for \textit{P. falciparum} from the case data set. The average household size in the study population is five and ranges from one to 20.

\begin{figure}[!ht]
\centering
\includegraphics[width=\textwidth]{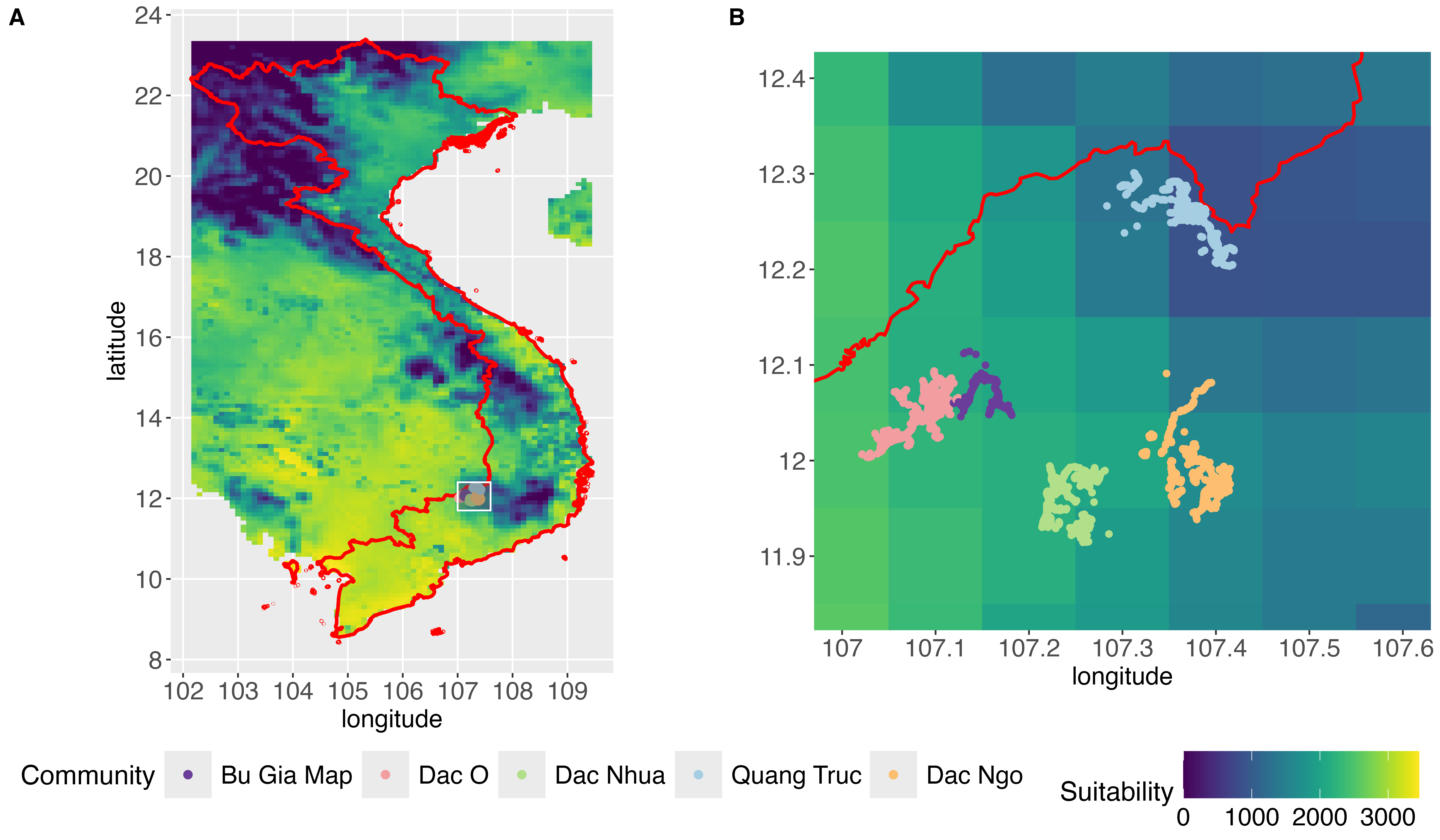}
\caption{Mosquito suitability in Vietnam (at a fixed time point, June) from Brown \textit{et al.} \cite{brown2024global}. The inset in Subplot A shows the region of interest. \textbf{B:} Mosquito suitability in the region of interest (also for June), showing the locations of the five communes in the case data.}
\label{fig:mosq_suit}
\end{figure}

\subsection{Mosquito suitability data}
We incorporate spatial differences in mosquito behaviour using the mosquito suitability maps by Brown \textit{et al.} \cite{brown2024global}, which provide an estimate of how suitable a location is for the survival of malaria mosquitoes and \textit{Plasmodium} parasites. An example of this suitability is presented in Figure \ref{fig:mosq_suit}A. For each month, there is a different suitability map that accounts for the impact of climatic variables on mosquitoes during that month. We will include malaria suitability estimates in the seasonally forced transmission within our mathematical model. Figure \ref{fig:mosq_suit}B is an inset of Figure \ref{fig:mosq_suit}A that illustrates the mosquito suitability map in the five study communes in Vietnam.\\

\section{Agent-Based Modelling}\label{ABM}
With the malaria case and mosquito suitability data in mind, we built an agent-based model that consists of human agents in households and mosquito agents with behaviour governed by suitability maps. 
\begin{figure}[!ht]
\centering
\includegraphics[width=\textwidth]{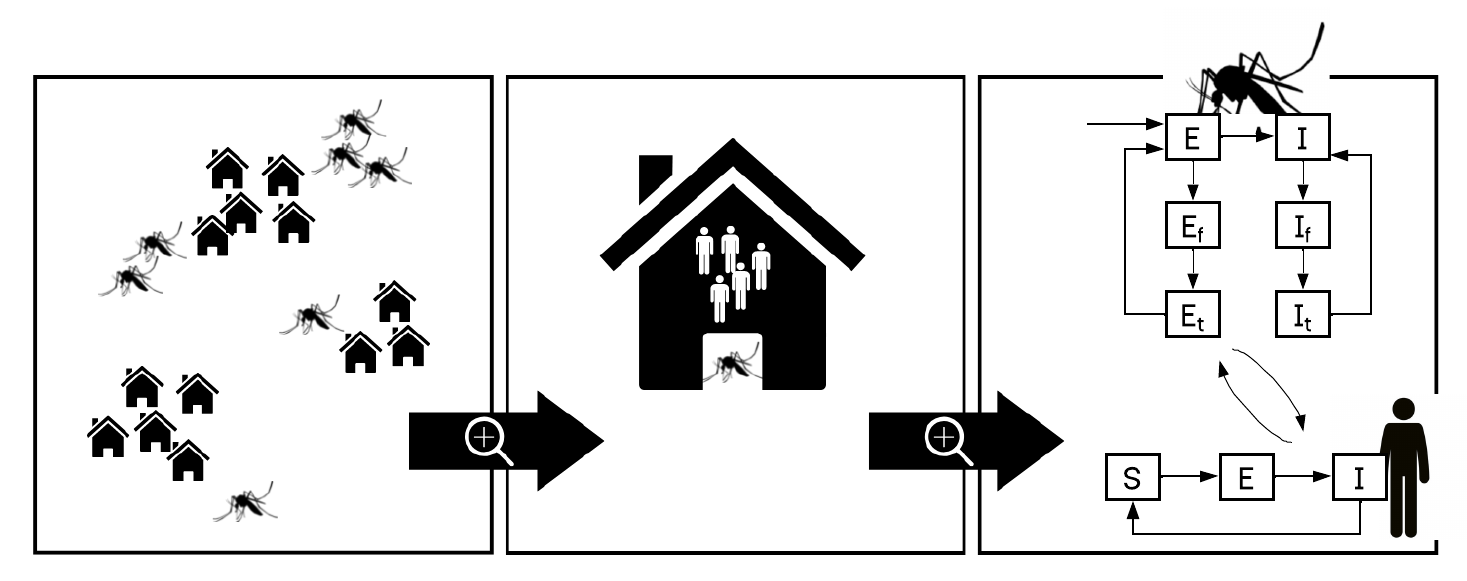}
\caption{Schematic representation of the ABM. Both humans and mosquitoes are modelled as agents, where human agents are localised in space, either in households grouped in villages. Mosquito agents are either localised in households or travelling between households. The disease progression in both populations is assumed to follow the Susceptible ($S$)-Exposed ($E$)-Infected ($I$) compartment paradigm. Within the mosquito population, exposed ($E$) and infectious ($E$) mosquitoes are further categorised as hungry, full or travelling.}
\label{fig:model}
\end{figure}
\subsection{Mosquito location map}

We assume that malaria suitability estimates have a 1-1 relationship with the frequency that mosquitoes visit a given location; however, the magnitude and scale of the relationship between malaria suitability and mosquito density are unknown. We let the value of the malaria suitability and mosquito density at household $h$ at time $t$ be denoted by $S_h(t)$ and $m(x_h,y_h,t)$ respectively, where $x_h\ \text{and }y_h$ are the longitude and latitude of household $h$, respectively. We define the relation between $m(x_h,y_h,t)$ and $S_h(t)$ as

\begin{table}[!ht]
\centering
\footnotesize
\caption{Definitions, values and sources for model parameters.}\label{tab:parameter}
\begin{tabular}[t]{clcc}
\toprule
\textbf{Parameter} & \textbf{Description} &  \textbf{Value} & \textbf{Source}  \\
\midrule
$b$ & Biting rate of mosquitoes & 1 day$^{-1}$& Assumed \\
$p_{HM}$ & Transmission probability: mosquito to human & 0.1& \cite{walker2023model} \\
$p_{MH}$ & Transmission probability: human to mosquito & 0.3& \cite{walker2023model}\\
$\beta_h(t)$& Rate at which mosquitoes are infected by household $h$& Varying \\
$r_{IRS}$ & Reduction in bites due to IRS & 0.3 & \cite{Scott:2017}  \\
$r_{LLINs}$ & Reduction in bites due to LLINs & 0.56 &  \cite{Scott:2017} \\
$d_{IRS}$ & Mosquito death rate due to IRS & 0.56 & \cite{Scott:2017} \\
$d_{LLINs}$ & Mosquito death rate due to LLIN & 0.19 & \cite{Scott:2017} \\
$1/\sigma_H$ & The mean exposed period in humans &9.9  day& \cite{Scott:2017} \\
$1/\sigma_M$& The mean exposed period in mosquitoes &14 day &\cite{Scott:2017} \\
$\gamma_M$ & Death rate of infectious mosquito &1/35 day$^{-1}$&\cite{Scott:2017}  \\
$\gamma_H$ & The natural recovery rate for human &1/30 day$^{-1}$&\cite{Scott:2017}  \\
$\eta$ & Rate of mosquitoes moving from full state to travelling state &1/2 day$^{-1}$&\cite{Scott:2017}  \\
$\alpha$ & Rate of mosquitoes travelling between households &5 km day$^{-1}$&\cite{Scott:2017}  \\
$p_{multi}$ & Probability that a mosquito takes multiple blood meals & 0.18& \cite{Norris:2010}\\
\bottomrule
\end{tabular}
\end{table}
\begin{equation}
m(x_h,y_h,t)=a S_h(t)^\theta, \label{suit_eqn}
\end{equation}
where $a$ and $\theta$ are scaling parameters. By varying $a$ and $\theta$, we can vary the intensity of transmission. Due to the low transmission in the study region, here we choose $a=1$ and $\theta=0.062$.

In our ABM, we do not model the mosquito bites that can not lead to an infection. Specifically, we focus on the bite rate of \textit{Anopheles} mosquitoes on humans already infected with malaria. We exclude mosquitoes that die from interventions, such as LLINs and IRS, after their first blood meal but before they can transmit the infection. In addition, the bite rate is adjusted based on the implementation of these interventions at the household level.

We model the number of bites of an individual in household $h$ as an inhomogeneous Poisson process with rate
\begin{equation}
\beta_h(t)= \frac{b p_{HM} I_h m(x_h,y_h,t)\left((1-r_{IRS})(1-d_{IRS})\right)^{IRS_h}\left((1-r_{LLIN})(1-d_{LLIN})\right)^{LLIN_h}}{N_h}, \label{mosq_gen}
\end{equation}
where $b$ is the average number of bites a mosquito makes when entering a home, $\beta_h(t)$ denotes the rate at which mosquitoes are infected by household $h$, $IRS_h$ is an indicator which takes the value 1 if IRS is applied to household $h$, $LLIN_h$ is an indicator which takes the value 1 if LLINs is applied to household $h$, $r_{IRS}$ is the reduction in bites due to IRS, $r_{LLIN}$ is the reduction in bites due to LLINs, $d_{IRS}$ is the probability of mosquitoes surviving to exit a household with IRS, $d_{LLIN}$ is the probability of mosquitoes surviving to exit a household with LLIN, $p_{HM}$ is the probability that an infected human infects a mosquito when bitten, $p_{MH}$ is the probability that a susceptible human gets infected from an infectious bite, $I_h$ is the number of infected individuals in household $h$ and $N_h$ is the total number of individual in household $h$.

For computational efficiency, we generate infected mosquitoes from each infected human at the rate

\begin{align}
    \frac{p_{HM}}{p_{MH}} \frac{m(x_h,y_h,t)\left((1-r_{IRS})(1-d_{IRS})\right)^{IRS_h}\left((1-r_{LLIN})(1-d_{LLIN})\right)^{LLIN_h}}{N_h}. \label{mosq_gen_rate}
\end{align}

This approach implicitly assumes that the overall mosquito population is large, resulting in negligible depletion of susceptible mosquitoes, and is therefore relatively constant within each month and across the region. Consequently, the mosquito bite rate is not affected by new infections (humans or mosquitoes).

\subsection{Human agents}
Individuals are localised in space, in households, and grouped in villages. Humans are susceptible in state $S$ and can be infected and move to state $E$ when bitten by an infectious mosquito. After an incubation period (average of $1/\sigma_H$) they move to state $I$ and can spread the disease to mosquitoes until they are treated (average time until treatment $1/\gamma_H$) and return to state $S$. Currently, we do not consider immunity against infection in our model.\\

With the bite rate $b$, each infectious bite causes an infection with probability $p_{MH}S_h/N_h$. Hence, an infectious, unfed mosquito in a household will cause infection at a rate 
\begin{align*}
    \hat{\beta}=\frac{b p_{MH} S_h}{N_h}.
\end{align*}

\subsection{Mosquito agents}
\subsubsection{Compartments}
Mosquitoes are infected according to the Poisson process given by Equation (\ref{mosq_gen}). Once infected, they are exposed on average for $1/\sigma_M$ (which is roughly 2 weeks \cite{Scott:2017}). In the exposed state ($E$), they can be exposed and hungry, exposed and full, or exposed and travelling (Figure \ref{fig:model}). After their exposed period, they become infectious ($I$); again, they can be hungry, exposed, or travelling. From any of these states, they can die naturally at a rate $\gamma_M$ (roughly a 30-day lifespan after becoming infected), or they can die from insecticides after biting a human with probability
$$1-(1-d_{IRS})^{IRS_h}(1-d_{LLIN})^{LLIN_h}.$$ They may become full when they bite. Although they may not become full, as there is evidence that up to 18\% will have blood meals from multiple humans \cite{Norris:2010}. They may also die before they are full. When full, they remain full for, on average, 2 days \cite{Scott:2017} and then travel to a new house. The travel time depends on the distance between the households (see Section \ref{mosq_movement}).

\subsubsection{Movement} \label{mosq_movement}
Let $d(i,j)$ denote the Euclidean distance between household $i$ and household $j$ and $\phi(a,b,c)$ denote the Gaussian probability density function with mean $b$ and standard deviation $c$ evaluated at $a$. We model mosquito movement between households, say $i$ and $j$, by assuming the mosquito chooses a destination (say household $j$ from household $i$) when it enters the travelling state at the rate \begin{equation*}
\frac{\eta m(x_j,y_j) \phi(d(i,j),0,\xi)}{\sum_{h\neq i} m(x_h,y_h)\phi(d(i,h),0,\xi)},
\end{equation*}
where $\eta$ is the rate at which a mosquito moves from the full state to the travelling state. Once the mosquito enters the travelling state, it will move towards the chosen household, say $j$, in a straight line at the rate
\begin{equation*}
\alpha/d(i,j),
\end{equation*}
where $\alpha$ is the rate of travel between households.

\subsection{Testing additional intervention strategies}
We investigate the impact of spatially distributed interventions on overall transmission dynamics. We consider two scenarios in which additional interventions are distributed to households that currently do not use any interventions ($n=2858$, Figure \ref{fig:all_data_hum}B): (i) distribution of a limited number of interventions and (ii) distribution of interventions with a limited budget. In the first scenario, we assume that resources are available to provide additional interventions to 1000 households (IRS only). In the second scenario, we assume that a limited budget of USD\$10,000 is available to provide additional interventions (IRS and/or LLINs) based on the best strategy in scenario (i). We take the cost associated with the distribution of IRS to a household is USD\$2.38 per household of size five and the cost associated with the distribution of LLINs is USD\$2.61 per person \cite{Scott:2017}. We investigated the impact of the additional intervention within one year in both scenarios under different strategies.\\

In identifying the houses to which IRS should be distributed, we consider factors, namely, mosquito suitability (Figure \ref{fig:all_data_hum}B), household size and a combination of both. Table \ref{tab:IRS} describes the different strategies adopted in scenario (i). Within strategy A, no additional interventions are distributed, that is, the baseline. Within strategy B, houses are ordered based on suitability and 1000 houses with the highest suitability are selected. Within strategies C and D, we chose 1000 households with the smallest and largest household size (members), respectively. To compare these selection strategies, we also randomly select 1000 households (strategy E). \\

In identifying the houses to which IRS and/or LLINs should be distributed with a fixed budget (here USD\$10,000), we consider houses with the smallest size as seen in scenario (i). Table \ref{tab:IRS_and_LLIN} describes the different types of distribution strategies adopted in Scenario (ii).\\

\begin{algorithm} 
\renewcommand{\thealgorithm}{1}
\caption{Agent-based model algorithm}\label{alg}
\begin{algorithmic}[1] 
\Statex \textbf{\textcolor{blue}{Initialisation:}}
\State Define model parameters and final simulation time, $T_{end}$. Read and process data.
\State Create a unique ID for each individual (person ID). Calculate the seasonal mosquito generation rate from each individual (Equation \eqref{mosq_gen_rate}).

\State Initialise $H_{stack}$ to track human transitions, household ID, and person ID and $ I_{stack}$ to track potential new human infection events, household ID, and person ID.\vspace{5mm}
\Statex \textbf{\textcolor{blue}{Generate infected mosquito from infected human:}}
\For{each initially infected human $i$}
    \State Record infection time, $t=t_I$ ($0$ for initially infected)
    \State \multiline{Generate recovery time, $t_R= t_I + \text{Exp} {(\gamma_H)}$ (exponentially distributed with rate $\gamma_H$). Update $H_{stack}$.}
    \State Propose the first time, $t_\Delta$ (Equation \eqref{mosq_gen_rate}), where human $i$ exposes a mosquito. 
    \While {$t+t_\Delta<\min{(t_R,T_{end})}$}
        \State \multiline{Set $t=t+t_\Delta$ and generate mosquito (exposed, full) into the household of human $i$.}
        \State \multiline{Generate time of becoming infectious, $t_M = t+\text{Exp} {(\sigma_M)}$, and time of death (natural), $t_D = t+\text{Exp} {(\gamma_M)}$. Set $t_{temp}=t$.}
        \While{$t_{temp}<t_D$ \& mosquito status is not `dead' \& $t_{temp}<T_{end}$}
            \If{mosquito status is `hungry'} it bites
                \State Generate time of bite,  $t_{temp}=t_{temp}+\text{Exp} {(b)}$
                
                  \If{hungry after a bite} 
                    \State keep status as `hungry' (it will bite again)
                  \ElsIf{full after a bite \& survives the IRS and LLINs}
                    \State update status to `full'
                    \Else
                    \State update status to `dead' (due to IRS and LLINs)
                  \EndIf
                  \If{$t_{temp}>t_M$ \& $t_{temp}< t_D$ \& $t_{temp}<T_{end}$}\vspace{5mm}
                  \Statex \textbf{\textcolor{blue}{Generate infected human from infectious mosquito:}}
                  \State \multiline{Since mosquito is alive and infectious at bite time, generate a possible human infection in the household. Update $I_{stack}$.}
                  \EndIf
            \ElsIf{mosquito status is `full'} it travels after a resting time
            \vspace{5mm}
                  \Statex \textbf{\textcolor{blue}{Mosquito travelling:}}
                \State Record time when ready to travel, $t_{temp}=t_{temp}+\text{Exp} {(\eta)}$.
                \State \multiline{Identify next household mosquito visits (based on distance,  suitability).}
            \Else \ (no longer `full' and travels)
                \State \multiline{Record time mosquito needs to travel to next household $t_{temp}=t_{temp}+\text{Exp} {(\alpha/\text{distance}(\text{current household, next household}))}$. Update mosquito location and update status to `hungry'.}
            \EndIf
        \EndWhile
        \State \multiline{Propose time at which next mosquito is exposed by infected human $i$, $t_{\Delta}$.}
    \EndWhile
\EndFor 
\State Set $T$ as the time of the first infection event in $I_{stack}$.

\algstore{myalg}
\end{algorithmic} 
\end{algorithm}

\begin{algorithm}[!h] 
\renewcommand{\thealgorithm}{1}
\begin{algorithmic} [1]                   
\caption{Agent-based model algorithm (continued)}
\algrestore{myalg} 
\Statex \textbf{\textcolor{blue}{Generate further infection unless time is over or no infectious humans:}}
\While{$T<T_{end}$ \& total infection events in $I_{stack}$ $\neq 0$}

\State Identify person with earliest possible human infection time in $I_{stack}$.
\If{the person is not in $H_{stack}$, or their last status in $H_{stack}$ is `recovered'}
\State \multiline{Person is susceptible at the time of infectious bite. Add person to $H_{stack}$, as exposed. Generate time they become infectious ($t_I$). }
\State Repeat Lines 8--33, with $i$ as ID of newly infected person.
\EndIf
\State \multiline{Remove earliest event from $I_{stack}$. Update $T$ to be the next infection time in $I_{stack}$.}
\EndWhile

\end{algorithmic} 
\end{algorithm}
\section{Results}
\subsection{Baseline disease burden}
\begin{figure}[!ht]
\centering
\includegraphics[width=\textwidth]{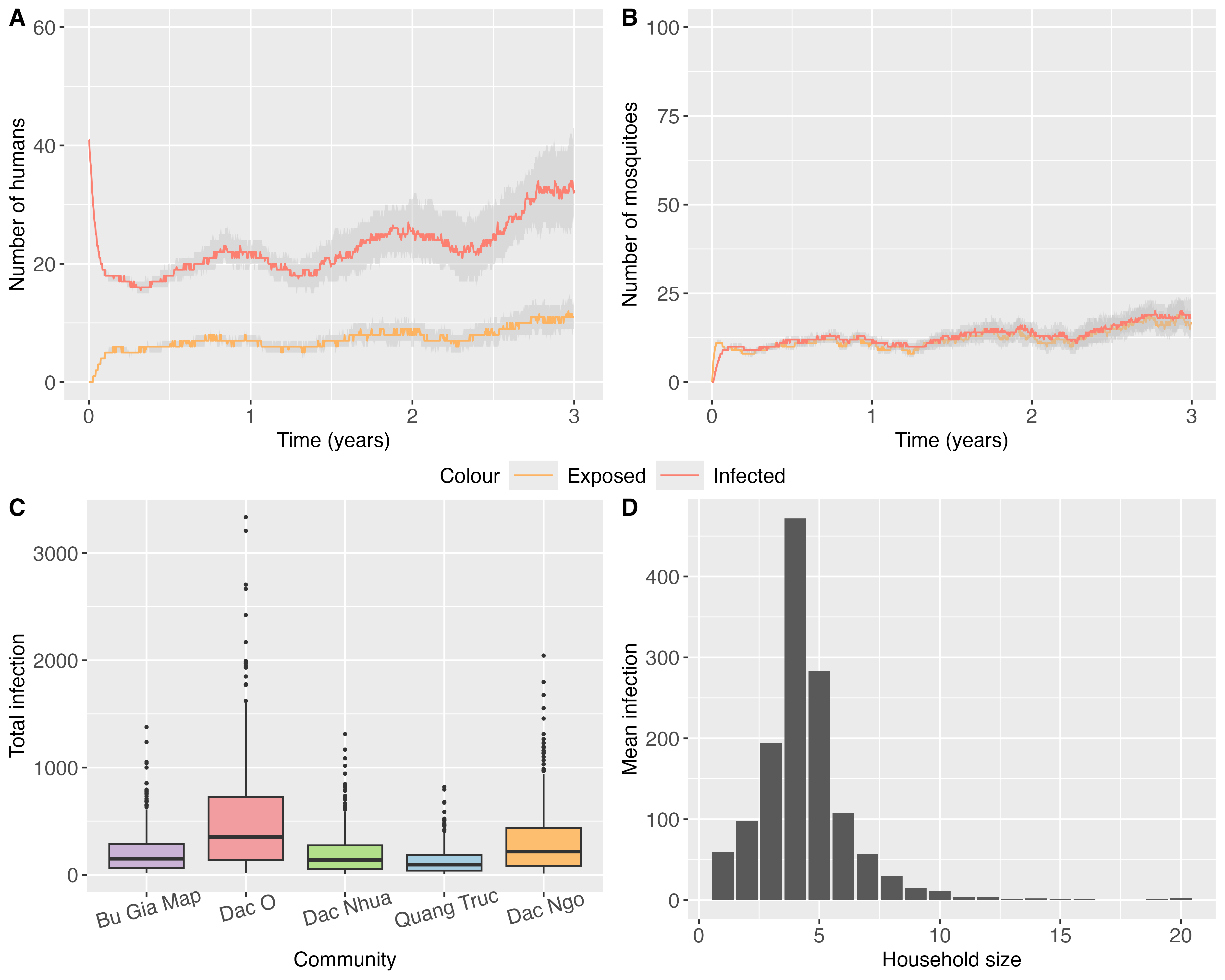}
\caption{Number of exposed and infected humans (Subplot A) and mosquitoes (Subplot B) under baseline treatments from 500 simulations. Subplots C and D depict the total number of infections per community and the mean number of infections by household size over the simulation period, respectively.}
\label{fig:hum_mosq}
\end{figure}

We use our spatio-temporal ABM to generate an approximation of the current disease burden in the region of interest. Figure \ref{fig:hum_mosq} shows a summary of the disease dynamics from 500 simulations in which no additional interventions were considered (baseline). The number of infected and exposed humans over time is depicted in Figure \ref{fig:hum_mosq}A, whereas Figure \ref{fig:hum_mosq}B depicts the number of exposed and infected mosquitoes. Figure \ref{fig:hum_mosq}C depicts the total number of infections across the five communities over three years, whereas the average total number of infections by household size is shown in Figure \ref{fig:hum_mosq}D.  With existing interventions in the region (baseline), our model indicates that the Dac O community has the highest malaria burden, followed by the Dac Ngo community (Figure \ref{fig:hum_mosq}C). The reason behind this is that the majority of the population (51.1\%), as well as the houses that currently do not use any intervention (85.2\%, Figure \ref{fig:all_data_hum}B), are located in the Dac O and Dac Ngo communities. Due to the median size of the households being four in the region (Figure \ref{fig:all_data_hum}C), the average infection is also the highest in these households (Figure \ref{fig:hum_mosq}D).\\
\begin{table}[H]
\centering
\footnotesize
\caption{Strategies for selecting 1000 houses (from those that currently use no intervention) to distribute IRS\label{tab:IRS}.}
\begin{tabular}[t]{cp{10cm}}
\toprule
\textbf{Strategy} & \textbf{Description}\\
\bottomrule
A& Baseline (no additional intervention)\\
B &Houses with the highest suitability\\
C& Houses with the smallest household size\\
D& Houses with the largest household size\\
E & Random houses\\
\bottomrule
\end{tabular}
\end{table}

\begin{table}[H]
\centering
\footnotesize
\caption{Strategies for distribution of IRS and/or LLINs with a budget of \$10,000\label{tab:IRS_and_LLIN}.}
\begin{tabular}[t]{cp{10cm}}
\toprule
\textbf{Strategy} & \textbf{Description}\\
\bottomrule
a& Baseline (no additional intervention)\\
 b&IRS+LLINs to houses that currently use none (based on smallest household size)\\
c&IRS (all houses that currently use none) and LLINs (houses that currently use none based on smallest household size)\\
 d&LLINs only (houses that currently use none based on smallest household size)\\
e &IRS (houses that currently use none) and IRS (houses with existing LLINs based on highest suitability)\\
\bottomrule
\end{tabular}
\end{table}

\subsection{Scenario (i)}
\begin{figure}[!ht]
\centering
\includegraphics[width=\textwidth]{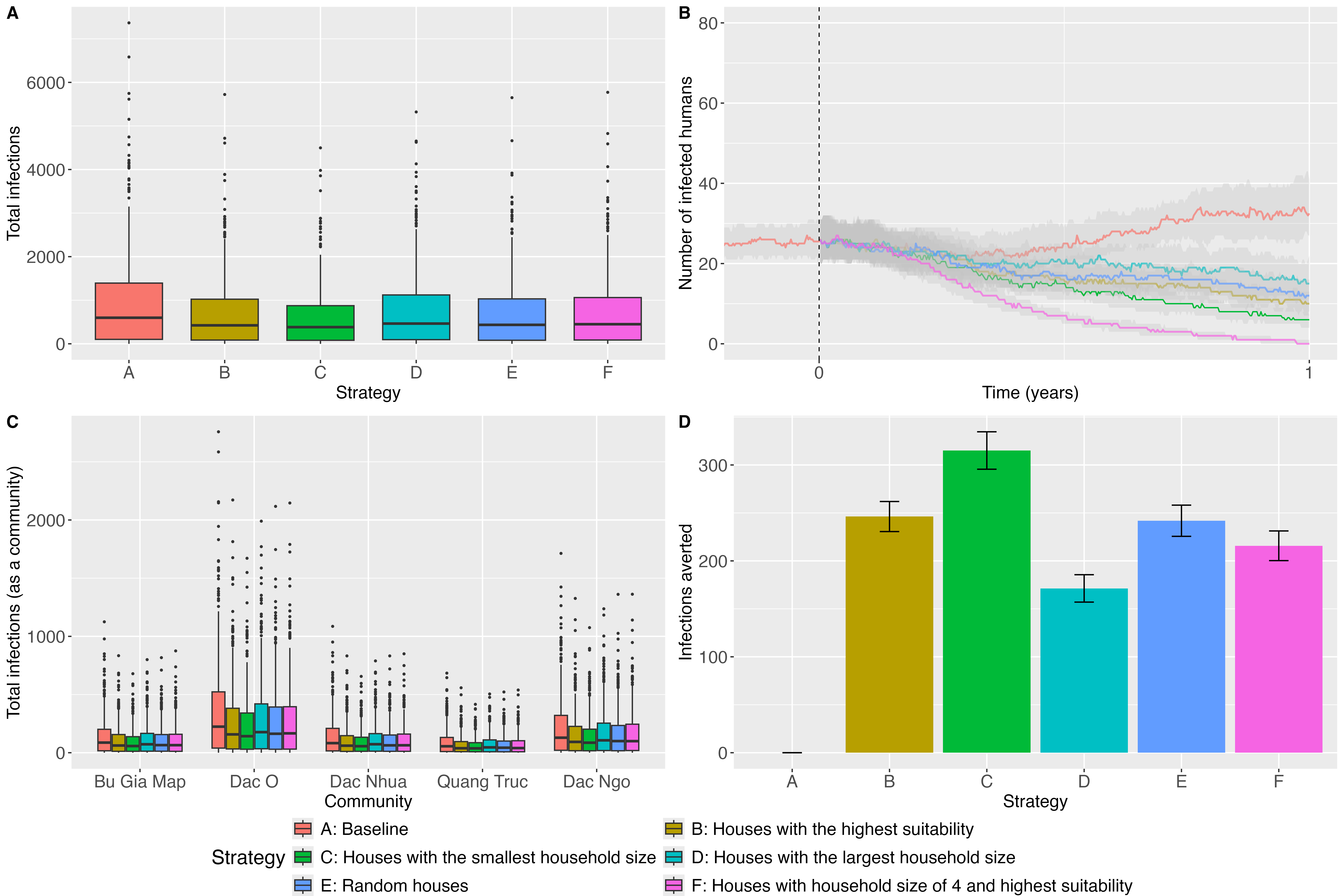}
\caption{Effect of distributing an additional 1000 IRS units to households that currently do not use any intervention via various strategies (Scenario (i)). Subplot A: distribution of the total number of infections in the region over a year. Subplot B: number of infected humans over time.  Subplot C: distribution of the total number of infections across the communities over a year. Subplot D: total number of infections averted with the additional intervention across different strategies, where the intervals show the standard error.}
\label{fig:scenario_i}
\end{figure}

In Scenario (i), we investigate the impact of additionally distributed interventions (1000 units of IRS) under different strategies of selecting houses. Figure \ref{fig:scenario_i} illustrates the summary of 500 simulations under Scenario (i). Figure \ref{fig:scenario_i}A and C show the total number of infections within a year in the region of interest and in each community using different strategies, respectively. The median number of infected humans and the associated 95\% prediction interval over time is depicted in Figure \ref{fig:scenario_i}B. The mean number of infections averted (with the standard error) across different strategies is depicted in Figure \ref{fig:scenario_i}D. The average infections in the Dac O and Dac Ngo communities are higher compared to other communities with or without additional interventions, since these two communities have the highest population size and households without intervention (Figure \ref{fig:all_data_hum}B). The median number of infected humans and the associated 95\% prediction interval over time is depicted in Figure \ref{fig:scenario_i}B, which demonstrates the effect of the additionally distributed IRS in reducing disease cases. The mean number of infections averted across different strategies is depicted in Figure \ref{fig:scenario_i}D with the standard error and clearly illustrates the impact of these different strategies for the selection of houses. The total number of infections averted is the highest with strategy C (houses with the smallest household size) and the lowest with strategy D (houses with the largest household size).\\

To understand why strategy C averts more infections than any of the other strategies, especially strategy D, in reducing the disease burden, note that the average number of mosquito bites an individual receives in a household depends on the household size (see Equation (\ref{mosq_gen})) as, the bites are divided among the members of the household. When a susceptible mosquito enters a household where at least one individual is infected, the probability of the mosquito being infected when taking a blood meal is $p_{HM}/N$. Hence, an infectious individual in a smaller household (without any interventions) is more likely (for example, $5\%$ with $N=2$) to infect a susceptible mosquito that enters the home compared to an individual in a larger household (for example, $1.25\%$ with $N=8$). Furthermore, given the probability of a mosquito making multiple bites in a household before being full or dead (1-$p_{full}$), individuals within a smaller household get bitten more frequently compared to those in larger households. As infectious mosquitoes are generated from infected individuals, the generation rate (Equation \eqref{mosq_gen_rate}) is higher for individuals within smaller households (Figure \ref{fig:HH112}B). Therefore, according to our model, protecting smaller households reduces the probability of infection (both for mosquitoes and humans) more than protecting larger households, ultimately reducing the overall burden of the disease. Figure \ref{fig:HH112}C illustrates the breakdown of the size of the households to which interventions are applied, by strategy. In strategy C, out of 1000 households, 797 households have a size less than the median household size of four (79.7\%) compared to 125 households in strategy D (12.5\%).\\

The total infections averted in strategy B (houses with the highest suitability) and strategy E (selecting houses randomly) are similar. When households are selected with strategy B, 69.3\% of the selected houses have a size less than the median size of the household, whereas in strategy E, 70\% of the selected houses have a household size less than the median size of the household(Figure \ref{fig:HH112}C). Figure \ref{fig:sup_HH112_com} illustrates the breakdown of the houses selected from the five communities in the strategies considered in scenario (i). Strategy B applied interventions only to households in Dac O, as the region with the highest mosquito suitability (Figure \ref{fig:all_data_hum}B and \ref{fig:priority}). In contrast, Strategy C applies interventions to households in the five communities. Hence, in combination, Strategy C is better in reducing the total infections compared to other strategies. Table \ref{tab:sum_tab} summarises the effect of the different strategies investigated in scenario (i), with the best identified strategy highlighted in green. 

\definecolor{lightgreen}{RGB}{222, 250, 210}
\begin{table}[!ht]
\centering
\footnotesize
\caption{Summary of the effect of different strategies for both scenarios. The shaded rows indicate the best strategies across the scenarios.\\}
\begin{NiceTabular}[t]{p{6cm}cc}\\
\toprule
\textbf{Strategies}    &\textbf{Total infection$\pm$ SE}    &\textbf{Infection averted$\pm$ SE}\\
\hline
\\
\multicolumn{3}{c}{\textbf{Scenario (i)}}\\
\hline
A: Baseline                                 &916$\pm$49              &0\\
B: Houses with the highest suitability       &670$\pm$36             &246$\pm$16\\

\rowcolor{lightgreen} C: Houses with the smallest household size   &601$\pm$31             &315$\pm$20\\
D: Houses with the largest household size     &745$\pm$39           &171$\pm$14\\
E: Random houses                               &674$\pm$35         &242$\pm$16\\
\hline
\\
\multicolumn{3}{c}{\textbf{Scenario (ii)}}\\
\hline
b: IRS+LLINs to houses that currently 
use none (based on smallest household size) &542$\pm$28         &374$\pm$22\\

\rowcolor{lightgreen}c: IRS (all houses that currently use none)
and LLINs (houses that currently use none 
based on smallest household size)           &458$\pm$23             &457$\pm$27\\
d: LLINs only (houses that currently 
use none based on smallest household size)  &584$\pm$30             &332$\pm$20\\
\rowcolor{lightgreen}e: IRS (houses that currently use none) 
and IRS (houses with existing LLINs based 
on highest suitability)                     &457$\pm$23             &459$\pm$27\\
\bottomrule
\end{NiceTabular}\label{tab:sum_tab}
\end{table}

\subsection{Scenario (ii)}
\begin{figure}[!ht]
\centering
\includegraphics[width=\textwidth]{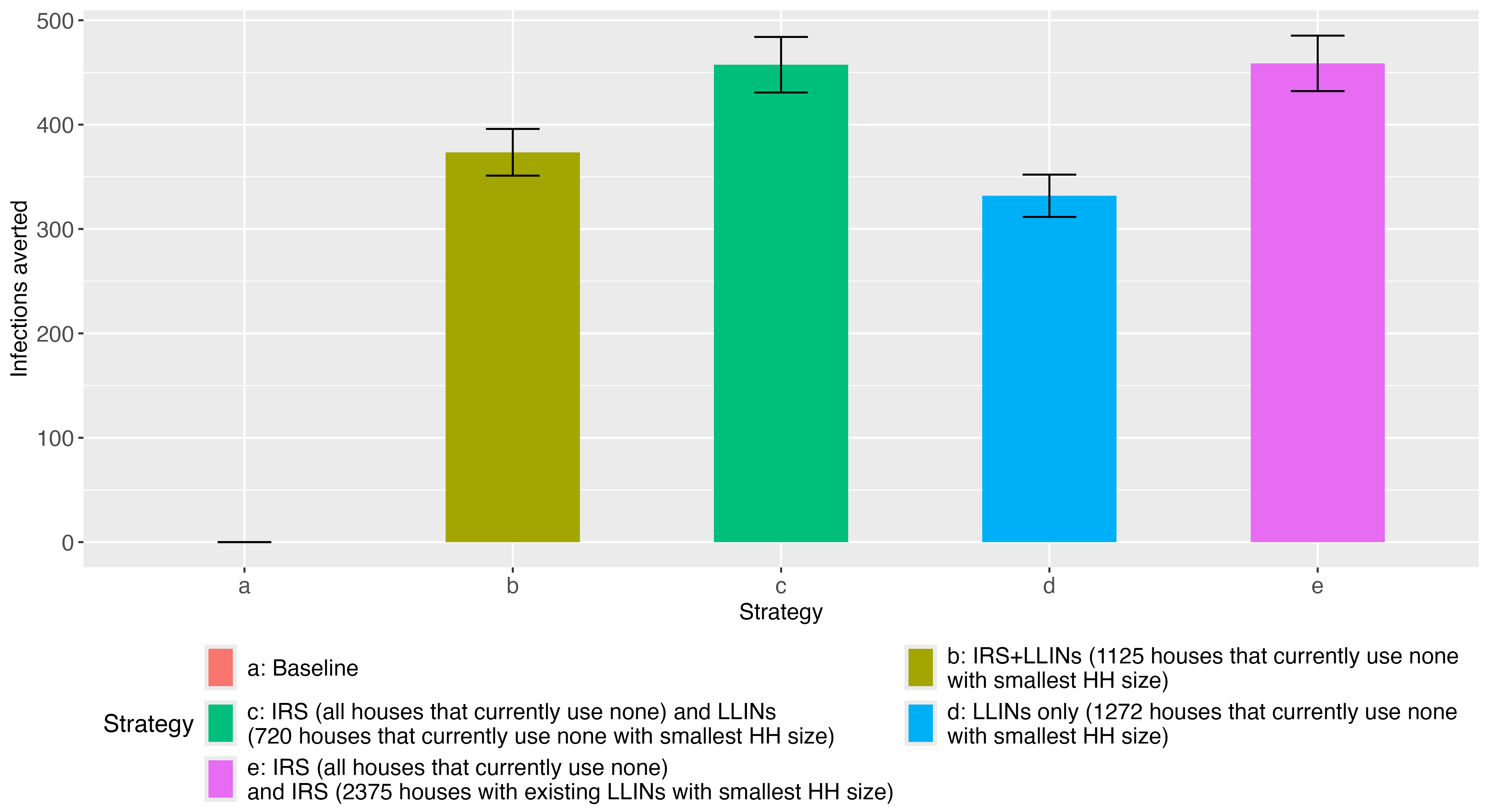}
\caption{Impact of distributing IRS and or LLINs to houses with a budget of USD\$10,000, showing the total number of infections averted under different selection strategies.}
\label{fig:Tot_avrtd_scenario2}
\end{figure}

Figure \ref{fig:Tot_avrtd_scenario2} illustrates the total number of infections averted under scenario (ii), where households are selected based on the strategies described in Table \ref{tab:IRS_and_LLIN}. With a fixed budget of USD\$10,000, providing some level of protection to a wide number of houses (strategy e in Figure \ref{fig:Tot_avrtd_scenario2}) seems to provide better results compared to strong protection provided to a smaller number of houses (strategy b, Figure \ref{fig:Tot_avrtd_scenario2}).  In strategy b, we estimate that IRS and LLINs can be provided to 1125 houses that currently do not use any intervention with the budget (out of 2858 households that do not use any interventions). Although the combination of IRS and LLINs provides better protection to households, strategy b is less effective than strategies that ensure the whole population is protected by at least one intervention. When IRS is distributed to all households that do not use any intervention, there is little difference between spending the remainder of the budget on LLINs for households with existing IRS (strategy c) or providing IRS to households with existing LLINs (strategy e). With just LLINs distributed to 1272 households under strategy d, it performs the worst in reducing the total infection burden. Table \ref{tab:sum_tab} summarises the effect of the different strategies investigated in both scenarios, with the best identified highlighted in green.

\section{Discussion}
Heterogeneity in susceptibility to infection is a crucial aspect of malaria transmission, especially in regions where seasonality is present in mosquito populations. Modelling the transmission dynamics with an ABM allows for the heterogeneity of individuals to be accounted for. In this work, we present a spatiotemporal agent-based malaria transmission model that captures the heterogeneity in mosquito bite exposure across households and applies it to a province in Vietnam. We use household data consisting of GPS location, household size, infected cases and current interventions in combination with mosquito suitability estimates \cite{brown2024global} to model the human and mosquito interaction in the region.\\ 

We investigated the impact of additional distributed interventions (either a fixed number of additional IRS units or with a fixed budget) using different household selection strategies on disease burden. According to our model, providing interventions to the smallest households that currently do not use any interventions provides the best results in reducing the burden of the disease with a fixed number of interventions (Figure \ref{fig:scenario_i}) compared to other strategies considered (Table \ref{tab:IRS}). Since we assume that the size of a household does not affect the number of mosquitoes that enter it and assume that mosquitoes are equally likely to bite any member of the household, it follows that the probability of a mosquito becoming infected is higher in smaller households than in larger ones (Figure \ref{fig:HH112}). Hence, reducing bites within smaller households by providing interventions reduces the intensity of transmission. In the case of a fixed budget, providing some level of protection to a larger number of households offers better protection to the entire region compared to a stronger level of protection to a small number of households (Figure \ref{fig:Tot_avrtd_scenario2}). These results reflect the assumptions of the model that the size of the household and the human infection status do not impact the behaviour of the mosquito. These results may vary if heterogeneity is considered within each household regarding exposure to mosquitoes, which in reality are affected by mosquito behaviour, individual body surface area, and hence gender and age \cite{gonccalves2017examining}. Hence, including demographic data on age and gender is an avenue for future work. \\

Agent-based models have been used to characterise the dynamics of malaria and the impact of treatments and interventions due to their ability to account for individual interactions, heterogeneity, and stochasticity \cite{Smith:2018}. In our model, we have not considered individual characteristics (e.g., immunity). Immunity plays a crucial role in susceptibility to infection, clinical risk and time to recovery in \textit{P. falciparum}, especially in high-transmission settings \cite{anwar2024mathematical}. Given that the study area is in a region of very low transmission, naturally acquired immunity is likely to play a small role in influencing transmission or clinical risk. The Vietnam province considered in this article was close to the border with Cambodia, which has a comparatively high malaria risk, especially from \textit{P. vivax}. Therefore, importation of cases can be an additional contributor to the burden of malaria and should be considered in future work. We currently only consider \textit{P. falciparum} in our model, however, \textit{P. vivax} poses a great challenge in malaria control efforts due to its ability to remain dormant and cause further infection when activated \cite{white2014modelling,anwar2024mathematical}. The presence of both species will challenge control efforts due to the need for both blood-stage and liver-stage treatments \cite{walker2023model}. Furthermore, \textit{P. falciparum} can mask a \textit{P. vivax} co-infection \cite{walker2023model,abba2011rapid,ashton2010performance}. Hence, including both \textit{P. falciparum} and  \textit{P. vivax} species in the model is an avenue for future potential work. Our model considers simple mosquito movement: when mosquitoes are in the travelling phase, they choose the next household based on its suitability, and then they travel directly towards that household. However, in reality, the mosquito may stop along the way and will not know the suitability at the destination in advance. In addition, the mosquito suitability to each household does not consider household size. However, larger houses may be more likely to attract more mosquitoes given an increase in $\text{CO}_2$ levels, as well as odours \cite{spitzen2016visualization,raji2017genetic}. We have not considered this in our model to avoid complexity. Furthermore, in order to choose an optimal strategy, other factors, such as situational awareness and distribution feasibility, should be incorporated into the ABM. 

Our spatial agent-based model provides a platform for investigating malaria control and elimination by accounting for spatial heterogeneity, which can be modified to apply to any geographical location \cite{brown2024global}. Our approach has the potential to aid malaria elimination efforts by investigating the impact of interventions on a small spatial scale. 

\section{Acknowledgements}
This research was supported by The University of Melbourne’s Research Computing Services and the Petascale Campus Initiative, Australia. This project was funded/supported by the Australian Centre of Research Excellence in Malaria Elimination (NHMRC 2024622). J.A. Flegg’s research is supported by the Australian Research Council (FT210100034) and the National Health and Medical Research Council (APP2019093). J. Richard's research is supported by the Regional Artemisinin Initiative (RAI) and the MRFF Career Development Fellowship GNT1161076. The funders had no role in the design, conduct, or analysis of the study.

\section{Data availability}
The household dataset used in this study contains sensitive information and cannot be shared publicly due to legal and ethical restrictions, including the need to protect individual privacy and comply with data sharing agreements with national health authorities. However, to give the reader an overview of how the ABM works, we have generated a synthetic dataset to mimic the real-world scenario. The model code, relevant datasets and supplementary video are available for the ABM in \href{https://github.com/n-anwar/Spatio_temporal_agent-based_model}{github} and provide an overview of how we have implemented Algorithm \ref{alg}.

\newpage
\printbibliography

@article{Huang:2011,
	author = {Huang, Fang and Zhou, Shuisen and Zhang, Shaosen and Wang, Hongju and Tang, Linhua},
	journal = {Malaria Journal},
	number = {1},
	pages = {1--8},
	publisher = {Springer},
	title = {Temporal correlation analysis between malaria and meteorological factors in Motuo County, Tibet},
	volume = {10},
	year = {2011}}

@article{Hay:2006,
	author = {Hay, Simon I and Snow, Robert W},
	journal = {PLoS Med},
	number = {12},
	pages = {e473},
	publisher = {Public Library of Science},
	title = {The Malaria Atlas Project: developing global maps of malaria risk},
	volume = {3},
	year = {2006}}

@article{Scott:2017,
	author = {Scott, Nick and Hussain, S Azfar and Martin-Hughes, Rowan and Fowkes, Freya JI and Kerr, Cliff C and Pearson, Ruth and Kedziora, David J and Killedar, Madhura and Stuart, Robyn M and Wilson, David P},
	journal = {Malaria journal},
	number = {1},
	pages = {1--14},
	publisher = {BioMed Central},
	title = {Maximizing the impact of malaria funding through allocative efficiency: using the right interventions in the right locations},
	volume = {16},
	year = {2017}}

@article{Endo:2018,
	author = {Endo, Noriko and Eltahir, Elfatih AB},
	journal = {GeoHealth},
	number = {3},
	pages = {104--115},
	publisher = {Wiley Online Library},
	title = {Environmental determinants of malaria transmission around the Koka Reservoir in Ethiopia},
	volume = {2},
	year = {2018}}

@article{Norris:2010,
	author = {Norris, Laura C and Fornadel, Christen M and Hung, Wei-Chien and Pineda, Fernando J and Norris, Douglas E},
	journal = {The American journal of tropical medicine and hygiene},
	number = {1},
	pages = {33--37},
	publisher = {ASTMH},
	title = {Frequency of multiple blood meals taken in a single gonotrophic cycle by Anopheles arabiensis mosquitoes in Macha, Zambia},
	volume = {83},
	year = {2010}}

@article{Smith:2018,
	author = {Smith, Neal R and Trauer, James M and Gambhir, Manoj and Richards, Jack S and Maude, Richard J and Keith, Jonathan M and Flegg, Jennifer A},
	journal = {Malaria Journal},
	number = {1},
	pages = {1--16},
	publisher = {BioMed Central},
	title = {Agent-based models of malaria transmission: a systematic review},
	volume = {17},
	year = {2018}}

@article{brown2024global,
  title={A global mathematical model of climatic suitability for Plasmodium falciparum malaria},
  author={Brown, Owen and Flegg, Jennifer A and Weiss, Daniel J and Golding, Nick},
  journal={Malaria Journal},
  volume={23},
  number={1},
  pages={306},
  year={2024},
  publisher={Springer}
}

@article{walker2023model,
	title = {A model for malaria treatment evaluation in the presence of multiple species},
 	author = {Walker, C. R. and Hickson, R. I. and Chang, E. and Ngor, P. and Sovannaroth, S. and Simpson, J. A. and Price, D. J. and {McCaw}, J. M. and Price, R. N. and Flegg, J. A. and Devine, A.},
	journal = {Epidemics},
	volume = {44},
 	pages = {100687},
  year={2023},
  publisher={Elsevier}
}

@article{gonccalves2017examining,
  title={Examining the human infectious reservoir for Plasmodium falciparum malaria in areas of differing transmission intensity},
  author={Gon{\c{c}}alves, Bronner P and Kapulu, Melissa C and Sawa, Patrick and Guelb{\'e}ogo, Wamdaogo M and Tiono, Alfred B and Grignard, Lynn and Stone, Will and Hellewell, Joel and Lanke, Kjerstin and Bastiaens, Guido JH and others},
  journal={Nature Communications},
  volume={8},
  number={1},
  pages={1133},
  year={2017},
  publisher={Nature Publishing Group UK London}
}

@article{white2014modelling,
  title={Modelling the contribution of the hypnozoite reservoir to Plasmodium vivax transmission},
  author={White, Michael T and Karl, Stephan and Battle, Katherine E and Hay, Simon I and Mueller, Ivo and Ghani, Azra C},
  journal={Elife},
  volume={3},
  pages={e04692},
  year={2014},
  publisher={eLife Sciences Publications, Ltd}
}

@article{anwar2024mathematical,
  title={Mathematical models of Plasmodium vivax transmission: A scoping review},
  author={Anwar, Md Nurul and Smith, Lauren and Devine, Angela and Mehra, Somya and Walker, Camelia R and Ivory, Elizabeth and Conway, Eamon and Mueller, Ivo and McCaw, James M and Flegg, Jennifer A and others},
  journal={PLOS Computational Biology},
  volume={20},
  number={3},
  pages={e1011931},
  year={2024},
  publisher={Public Library of Science San Francisco, CA USA}
}

@article{unicef2022fighting,
  title={Fighting malaria with long-lasting insecticidal nets (LLINs)},
  author={Unicef and others},
  year={2022}
}

@article{santos2021unsustainability,
  title={The Unsustainability of Long-Lasting Insecticidal Nets},
  author={Santos, Ellen M and Curtis, Tatyana M},
  journal={The American Journal of Tropical Medicine and Hygiene},
  volume={105},
  number={4},
  pages={876},
  year={2021},
  publisher={The American Society of Tropical Medicine and Hygiene}
}

@article{pluess2010indoor,
  title={Indoor residual spraying for preventing malaria},
  author={Pluess, Bianca and Tanser, Frank C and Lengeler, Christian and Sharp, Brian L},
  journal={Cochrane Database of Systematic Reviews},
  number={4},
  year={2010},
  publisher={John Wiley \& Sons, Ltd}
}

@article{zhou2022effectiveness,
  title={Effectiveness of indoor residual spraying on malaria control: a systematic review and meta-analysis},
  author={Zhou, Yiguo and Zhang, Wan-Xue and Tembo, Elijah and Xie, Ming-Zhu and Zhang, Shan-Shan and Wang, Xin-Rui and Wei, Ting-Ting and Feng, Xin and Zhang, Yi-Lin and Du, Juan and others},
  journal={Infectious Diseases of Poverty},
  volume={11},
  number={04},
  pages={29--42},
  year={2022},
  publisher={Editorial Office of Infectious Diseases of Poverty, National Institute of~…}
}

@incollection{riveron2018insecticide,
  title={Insecticide resistance in malaria vectors: an update at a global scale},
  author={Riveron, Jacob M and Tchouakui, Magellan and Mugenzi, Leon and Menze, Benjamin D and Chiang, Mu-Chun and Wondji, Charles S},
  booktitle={Towards Malaria Elimination-a Leap Forward},
  year={2018},
  publisher={IntechOpen}
}

@book{world2024world,
  title={World malaria report 2024},
  author={World Health Organization},
  year={2024},
  publisher={World Health Organization}
}

@book{keeling2011modeling,
  title={Modeling infectious diseases in humans and animals},
  author={Keeling, Matt J and Rohani, Pejman},
  year={2011},
  publisher={Princeton University Press}
}

@article{huppert2013mathematical,
  title={Mathematical modelling and prediction in infectious disease epidemiology},
  author={Huppert, Amit and Katriel, Guy},
  journal={Clinical Microbiology and Infection},
  volume={19},
  number={11},
  pages={999--1005},
  year={2013},
  publisher={Elsevier}
}

@article{abba2011rapid,
  title={Rapid diagnostic tests for diagnosing uncomplicated P. falciparum malaria in endemic countries},
  author={Abba, Katharine and Deeks, Jonathan J and Olliaro, Piero L and Naing, Cho-Min and Jackson, Sally M and Takwoingi, Yemisi and Donegan, Sarah and Garner, Paul},
  journal={Cochrane database of systematic reviews},
  number={7},
  year={2011},
  publisher={John Wiley \& Sons, Ltd}
}

@article{ashton2010performance,
  title={Performance of three multi-species rapid diagnostic tests for diagnosis of Plasmodium falciparum and Plasmodium vivax malaria in Oromia Regional State, Ethiopia},
  author={Ashton, Ruth A and Kefyalew, Takele and Tesfaye, Gezahegn and Counihan, Helen and Yadeta, Damtew and Cundill, Bonnie and Reithinger, Richard and Kolaczinski, Jan H},
  journal={Malaria Journal},
  volume={9},
  pages={1--11},
  year={2010},
  publisher={Springer}
}

@article{khanh2025unprecedented,
  title={Unprecedented large outbreak of Plasmodium malariae malaria in Vietnam: Epidemiological and clinical perspectives},
  author={Khanh, Chau Van and L{\^e}, Hương Giang and V{\~o}, Tuấn Cường and Quang, Nguyen Xuan and Nguyen, Do Van and Dung, Nguyen Cong Trung and Tam, Le Thanh and Nhien, Nguyen Thanh Thuy and Nguyễn, {\DJ}{\u{a}}ng Th{\`u}y Dương and Nguyễn, Thu Hằng and others},
  journal={Emerging Microbes \& Infections},
  volume={14},
  number={1},
  pages={2432359},
  year={2025},
  publisher={Taylor \& Francis}
}

@article{bannister2018micro,
  title={Micro-epidemiology of malaria in an elimination setting in Central Vietnam},
  author={Bannister-Tyrrell, Melanie and Xa, Nguyen Xuan and Kattenberg, Johanna Helena and Van Van, Nguyen and Dung, Vu Khac Anh and Hieu, Truong Minh and Van Hong, Nguyen and Rovira-Vallbona, Eduard and Thao, Nguyen Thanh and Duong, Tran Thanh and others},
  journal={Malaria Journal},
  volume={17},
  pages={1--15},
  year={2018},
  publisher={Springer}
}

@article{van2023anopheles,
  title={Anopheles diversity, biting behaviour and transmission potential in forest and farm environments of Gia Lai province, Vietnam},
  author={Van Dung, Nguyen and Thieu, Nguyen Quang and Canh, Hoang Dinh and Le Duy, Bui and Hung, Vu Viet and Ngoc, Nguyen Thi Hong and Mai, Nguyen Vu Tuyet and Van Anh, Ngo Thi and Son, Le Duy and Oo, Win Han and others},
  journal={Malaria Journal},
  volume={22},
  number={1},
  pages={204},
  year={2023},
  publisher={Springer}
}

@article{spitzen2016visualization,
  title={Visualization of house-entry behaviour of malaria mosquitoes},
  author={Spitzen, Jeroen and Koelewijn, Teun and Mukabana, W Richard and Takken, Willem},
  journal={Malaria journal},
  volume={15},
  pages={1--10},
  year={2016},
  publisher={Springer}
}

@article{raji2017genetic,
  title={Genetic analysis of mosquito detection of humans},
  author={Raji, Joshua I and DeGennaro, Matthew},
  journal={Current opinion in insect science},
  volume={20},
  pages={34--38},
  year={2017},
  publisher={Elsevier}
}

\newpage
\appendix
\section{Supplementary Figures}

\setcounter{figure}{0}
\renewcommand{\thefigure}{S\arabic{figure}}

\begin{figure}[!ht]
\centering
\includegraphics[width=\textwidth]{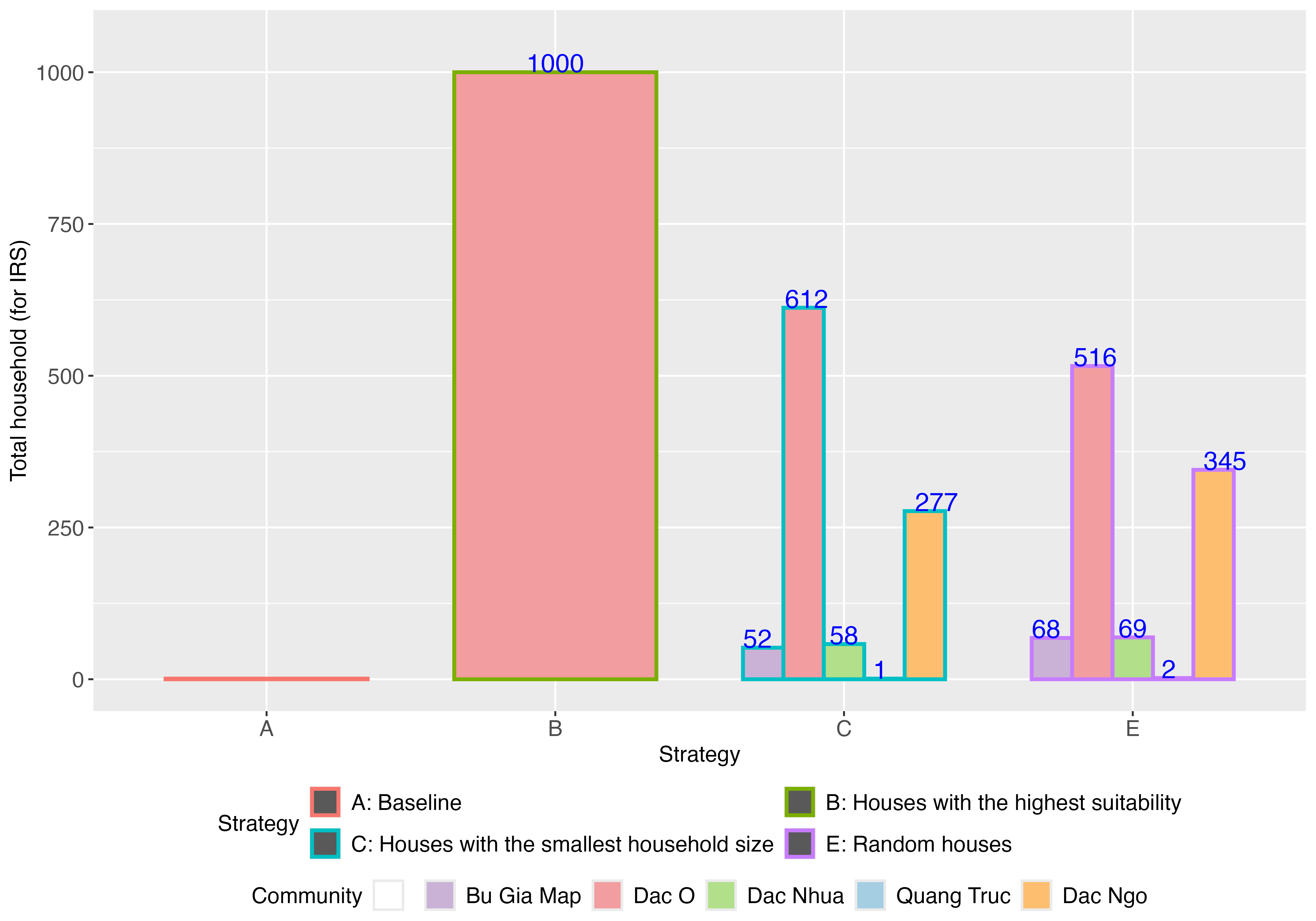}
\caption{Selected houses across all communities for additional intervention in different household selection scenarios (Scenario (i)).}
\label{fig:sup_HH112_com}
\end{figure}

\begin{figure}[!ht]
\centering
\includegraphics[width=\textwidth]{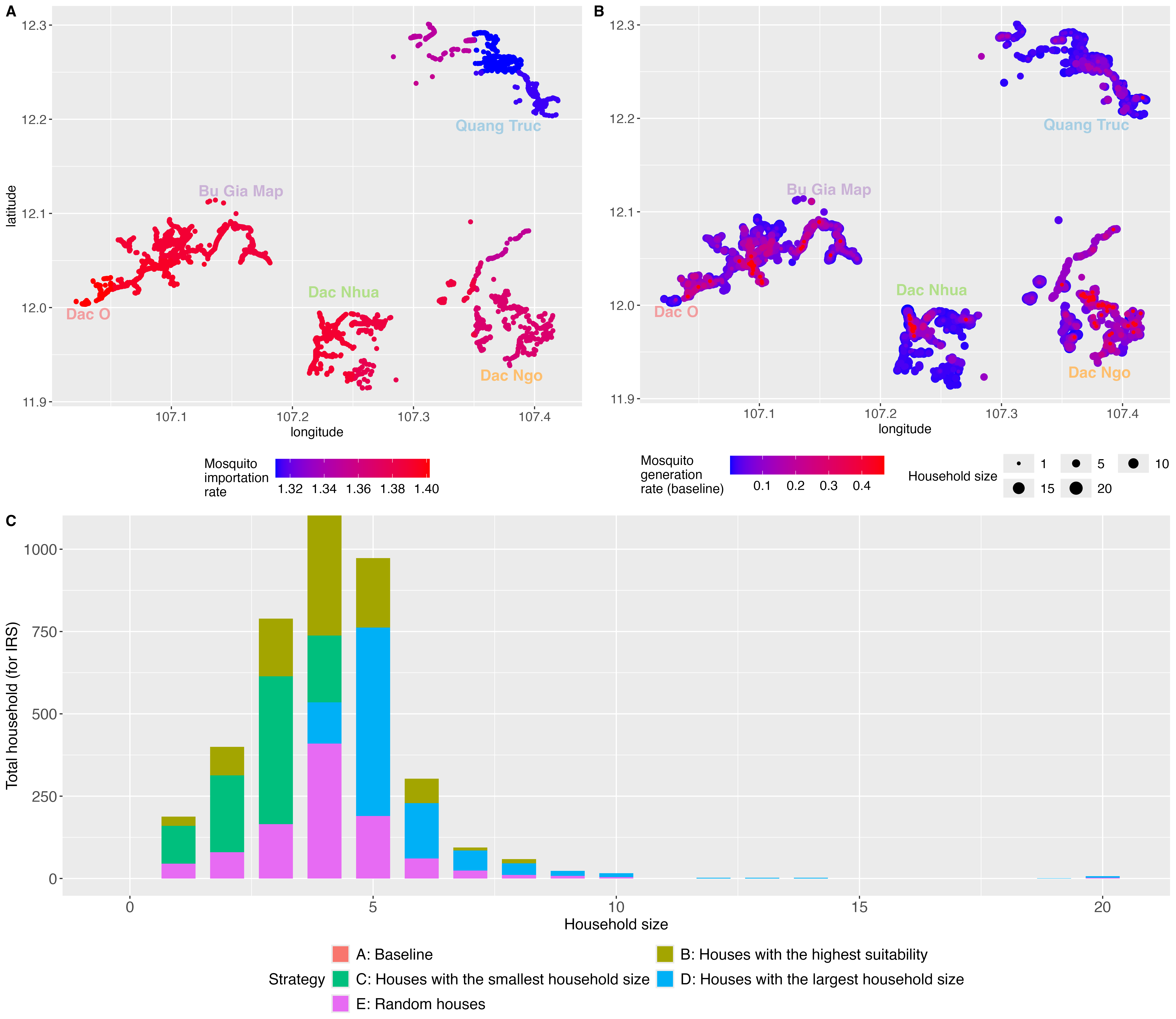}
\caption{Mosquito importation rate per household before any interventions (Subplot A) and mosquito generation rate with baseline interventions (Subplot B). Household size within selected houses for additional intervention in different intervention scenarios (Subplot C) (Scenario (i)).}
\label{fig:HH112}
\end{figure}

\begin{figure}[!ht]
\centering
\includegraphics[width=\textwidth]{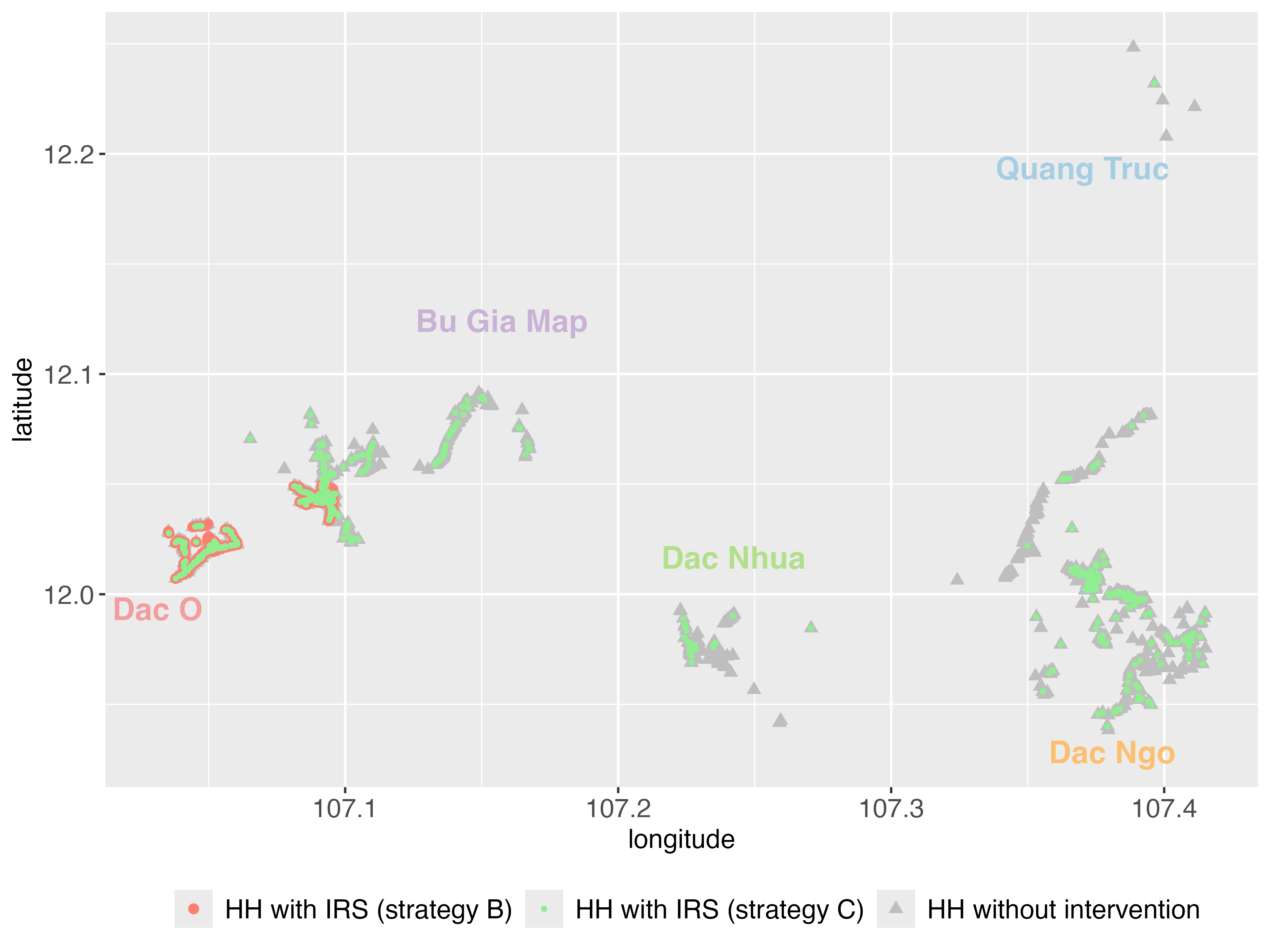}
\caption{Distribution of IRS to houses without any intervention based on strategy B (highest suitability) and Strategy C (smallest household size) in Scenario (i).}
\label{fig:priority}
\end{figure}

\end{document}